\documentclass[11pt]{article}
\usepackage{moriond,epsfig}

\def\be{\begin{equation}}
\def\ee{\end{equation}}
\def\bea{\begin{eqnarray}}
\def\eea{\end{eqnarray}}

\input rmpsym

\begin{document}
\vspace*{4cm}
\title{\CP\ VIOLATION: RECENT RESULTS FROM \babar}

\author{ G. Hamel de Monchenault\footnote[1]{e-mail address: {\tt gautier@hep.saclay.cea.fr}},
\\ on behalf of the \babar\ Collaboration }
\address{DSM/Dapnia, CEA/Saclay, \\
 F-91191 Gif-sur-Yvette, France }

\maketitle\abstracts{ We review recent time-dependent measurements
in the \B\ meson sector based on data collected between 1999 and
2002 by the \babar\ detector, which correspond to approximately 88
millions \BB\ pairs.   }

\section{Introduction}

The striking agreement with the Standard Model  predicted
value~\cite{ckm_fitters} of direct measurements of \stwob\ at \B\
Factories, by \babar~\cite{babar_stwob}
\begin{eqnarray}
\stwob &=& 0.741 \pm 0.067\,({\rm stat}) \pm 0.034\,({\rm syst})
\end{eqnarray}
and \belle~\cite{belle_stwob} experiments, leaves little room for
sizable New Physics contribution to the \BzBzb\ flavor
mixing~\cite{nir}. Because \stwob\ is turning to a precision
measurement,  \babar\ has started to test experimentally some of
the basic assumptions in the interpretation of the time-dependent
\CP\ asymmetry in "Golden" charmonium modes in terms of
\stwob~\cite{babar_gautier}: constraints on direct \CP\ violation
in $B \to \jpsi K$ decays; limit on wrong-flavor transitions in $B
\to \jpsi K$ decays; test of \CPT\ conservation and limits on \CP\
and \T\ non-conservation in \BzBzb\ mixing; and limit on the
lifetime difference between neutral \B\ mesons. In addition are
performed measurements of \stwob\ using various neutral \B\ decay
modes involving different short-distance physics: $\phi \KS$,
$\eta^\prime \KS$, $D^{\star \pm}D^\mp$, $D^{\star +}D^{\star -}$
and $\jpsi \pi^0$; and time-dependent \CP\ asymmetry measurements
in charmless \B\ decay modes, $\pi^+\pi^-$ and $\rho^\pm \pi^\mp$,
which are related to angle $\alpha$ of the Unitarity Triangle.

\subsection{The data set}
Unless specified otherwise, the measurements presented in this
paper are based on a data sample of about 88 million \BB\ pairs
collected between 1999 and 2002 with the \babar\ detector at the
\pepii\ asymmetric-energy \BF\ at SLAC. This corresponds to an
integrated luminosity of approximately 92\invfb\ at the \FourS\
resonance.  We also exploit a sample of 9\invfb\ of data taken
40\mev\ below the resonance (off-resonance data) for continuum
background studies.

\subsection{The \babar\ detector} The \babar\ detector is described
in detail elsewhere~\cite{babar_nim}. The tracking system is
composed of a cylindrical drift chamber (DCH) and a silicon vertex
tracker (SVT), both operating in a 1.5-T solenoidal magnetic
field. Charged particle identification is performed using a
ring-imaging Cherenkov detector (DIRC) and ${\rm d}E/{\rm d}x$
information from tracking detectors. Electrons and photons are
identified and their energy measured with a CsI electromagnetic
calorimeter (EMC). Muons and neutral hadrons are identified in the
instrumented flux return (IFR).

\subsection{The \BzBzb\ system}
There are three relevant bases to describe the neutral \B\ meson
system. The first basis is that of the two flavor eigenstates \Bz\
and \Bzb, which are related through \CP\ transformation. The
second basis is that of the \CP\ eigenstates of the system, $B_+$
and $B_-$.  The third basis is that of the physical states that
propagate with definite mass and lifetime. The mass difference and
width difference between the "heavy" $B_H$ and "light" $B_L$
mesons are defined as
\begin{eqnarray}
  \Delta m \equiv m_{B_H}-m_{B_L}>0\, , \ \ \Delta \Gamma \equiv {
  \Gamma}_{B_H} - {\rm \Gamma}_{B_L} \ . \nonumber
\end{eqnarray}
$\Delta m$, which is $2 \pi$ times the \BzBzb\ flavor oscillation
frequency, is measured with great accuracy. The present world
average,  $\Delta m = 0.502 \pm 0.006\ {\rm ps}^{-1}$, is
dominated by measurements at \B\ Factories~\cite{hfag}. A
peculiarity of the \BzBzb\ system is that the oscillation
frequency and the width are of the same order,  $x_d = \Delta m /
\Gamma = 0.755 \pm 0.015$. Only very loose constraints of the
width difference, $\Delta \Gamma$, are available up to
now~\cite{dG_constraints}. In the Standard Model, $\Delta \Gamma /
\Gamma$ is proportional to $m_b^2 / m_t^2$ and thus is expected to
be very small.  A recent leading order calculation yields $\Delta
\Gamma / \Gamma = -0.3\%$~\cite{dighe}, but next-to-leading order
corrections are expected to be large and may lead to an even
smaller absolute value~\cite{becirevic}.

\subsection{ \CP, \T\ and \CPT\ conservation in \BzBzb\ mixing alone}

The discrete symmetries \CP\ and \T\ are expected to be violated
in \BzBzb\ mixing alone, but at a very low level.  \CP\ and \T\
violation can be characterized by the parameter $q / p$, where $q$
and $p$ are (assuming \CPT\ invariance) the reduced complex
coefficients that link the mass and flavor eigenstates of the
system according to
\begin{eqnarray}\label{eq:physicalStates}
  |B_L\rangle = p\, |\Bz\rangle + q\, |\Bzb\rangle,\ \ \
  |B_H\rangle = p\, |\Bz\rangle - q\, |\Bzb\rangle \, .  \nonumber
\end{eqnarray}
The argument of $q / p$ depends on phase conventions; therefore $q
 / p$ is not an observable, but its modulus $|q / p|$ is.  A
departure of $|q / p|$ from $1$ is a manifestation of both \CP\
and \T\ violation in \BzBzb\ mixing alone.  In the Standard Model
this effect is tiny: $|q / p| - 1 \simeq 4 \pi  \m_c^2 / m_t^2 \
\sin {\beta} \approx 5 \times 10^{-4}$.  For most applications one
can write $q / p = e^{2i\phi_M}$, where $\phi_M= -\beta$ in the
usual Wolfenstein phase convention~\cite{PDG}.
\par
\CPT\ conservation, based on very general principles of
relativistic quantum mechanics, relies on the locality of quantum
field theories. However some theories in modern physics, such as
string theories, are not local at very short distances. Therefore
the \CPT\ symmetry could be violated. We introduce the phase
convention-independent complex parameter $z \equiv ( \delta M - (i
/ 2 ) \delta \Gamma ) /  ( \Delta m - (i / 2 ) \Delta \Gamma$ ),
where $ \delta M $ and $\delta \Gamma$ are differences between the
diagonal elements of the mass (dispersive) and lifetime
(absorptive) components of the effective Hamiltonian describing
the evolution of the neutral \B\ meson system.  $z\neq 0$ is a
manifestation of both \CP\ and \CPT\ violation in \BzBzb\ mixing
alone.

\subsection{Time evolution at the \FourS}
\B\ Factories are energy-asymmetric \epem\ colliders operating at
an energy of $\sqrt{s} = 10.58 \gev$ on the \FourS\ resonance. The
\FourS\ is the $4^{3}S_{1}$ state ($J^{PC}=1^{--}$) of the
bottonium ($b \bar{b}$) system; it decays exclusively into a \BB\
pair, $B^+B^-$ or \BzBzb\ in nearly equal amounts. The \BzBzb\
system from a \FourS\ decay evolves coherently:  the two mesons
flavor-oscillate in phase in such a way that at any moment in
time, the system is the superposition of exactly one \Bz\ and one
\Bzb\ meson. The decay of one meson serves as an analyzer of the
state of the accompanying meson at that instant.
\par
Experimentally, we reconstruct fully the decay on one of the two
\B\ mesons, that we label $B_{\rm rec}$, into a final state
$f_{\rm rec}$.  The other particles, which form the {\it rest of
the event} (ROE), come from the decay of the second \B\ meson,
$B_{\rm tag}$, into the final state $f_{\rm tag}$. The odds of
reconstructing fully the final state $f_{\rm tag}$ are small.
Instead, we apply a flavor tagging algorithm to the ROE based on
the presence of charged lepton, kaons, soft pions, and other
kinematical properties.  This algorithm determines the flavor of
the $B_{\rm tag}$ with an effective efficiency ({\it i. e.} taking
into account the probability of wrong flavor assignment) of
$28.1\pm0.7\%$.  The sample of selected events is divided
according to the result of the flavor tagging algorithm, either
\Bz\ or \Bzb\ tag. The proper time difference $\Delta t$ between
the two decays is deduced from the measurement of the distance
$\Delta z$ between the $B_{\rm rec}$ and $B_{\rm tag}$ decay
vertices along the boost axis, which is measured with a resolution
(RMS) of about $150 \mum$.  (At \pepii, the distance between the
two $B$ vertices is $260 \mum$ in average.)
\par
We consider cases where $f_{\rm rec}$ is a \CP\ eigenstate
$f_{\CP}$, {\it i.e.} such that  $\CP | f_{\CP} \rangle =
\eta_{f_{\CP}} | f_{\CP} \rangle$ with $\eta_{f_{\CP}} = \pm 1$.
We introduce a convention-independent complex parameter
\begin{eqnarray}
  \lambda_{f_{\CP}} \equiv \frac{q}{p} \frac{
  \overline{A}_{f_{\CP}}}{ {A}_{f_{\CP}}} ,     \nonumber
\end{eqnarray}
where we use the notation  $ {\kern 0.18em\optbar{\kern -0.18em
A}{}}_f \equiv \mathcal{A} (\, \BorBbar \to f \, )$ for the
complex decay amplitudes. The imaginary part of $
\lambda_{f_{\CP}}$ characterizes \CP\ violation in the
interference between \BzBzb\ mixing and \Bz\ or \Bzb\ decay while
a value of $|\lambda_{f_{\CP}}|$ different from 1 is an indication
of direct \CP\ violation in the decay (assuming that the effects
of \CP\ violation in \BzBzb\ mixing alone are negligible).   The
time-dependent \CP\ asymmetry
\begin{eqnarray}
  a_{f_{\CP}}(\Delta t) \equiv  \frac{ \mathcal{N}( \Delta t; \Bz\, {
  \rm tag} ) -  \mathcal{N}( \Delta t; \Bzb\, {
  \rm tag} ) } { \mathcal{N}( \Delta t; \Bz\, {
  \rm tag} ) +  \mathcal{N}( \Delta t; \Bzb\, {
  \rm tag} ) }    \nonumber
\end{eqnarray}
can be expressed (omitting effects of imperfect flavor tagging and
time difference reconstruction) as
\begin{eqnarray}
  a_{f_{\CP}}(\Delta t)  = -C_{f_{\CP}}\, \cos{ ( \Delta m \, \Delta t ) } +
  S_{f_{\CP}}\, \sin{ ( \Delta m \, \Delta t ) } \nonumber \\ \ \ {\rm with} \
  \ C_{f_{\CP}} \equiv \frac{ 1-|\lambda_{f_{\CP}}|^2 } { 1+|\lambda_{f_{\CP}}|^2 } \ \
  {\rm and} \ \ S_{f_{\CP}} \equiv \frac{ 2\, {\rm Im}\, \lambda_{f_{\CP}} } { 1+|\lambda_{f_{\CP}}|^2
  }\nonumber \ .
\end{eqnarray}
\par
We also consider cases where $f_{\rm rec}$ is a flavor eigenstate,
or very nearly so, {\it i.e.} such that $|\overline{A}_f| \ll
|A_f|$. This larger sample is used to determine from the data the
probabilities of wrong flavor-tag assignment and the parameters of
the time resolution function.

\section{Limit on direct \CP\ violation in $\jpsi K$ decays}

In the Standard Model \CP\ violation is expected to be small in $B
\to \jpsi K$ decays.  The reason is that the sub-dominant
contribution to the decay, a gluonic penguin amplitude, has the
same weak phase as the dominant color-suppressed tree amplitude.
The first contribution to the decay with a different phase is
highly suppressed.  One expects $| \overline{A}_{\jpsi \KS} /
A_{\jpsi \KS} | -1 \leq 10^{-2}$.

The test of the absence of direct \CP\ violation in these decays
is two-fold.  On the $\CP=-1$ sample used for the measurement of
\stwob\ we obtain~\cite{babar_stwob}:
\begin{eqnarray}
|\lambda_{\jpsi \KS}| = 0.948 \pm 0.051 \pm 0.030 \ .
\end{eqnarray}
We also measure the charge asymmetry for the isospin-related
charged mode. Based on a sample of about $1300$ $\jpsi K^{\pm}$
candidates selected in 20.7\invfb\ of data, we
find~\cite{babar_jpsik}:
\begin{eqnarray}
{\mathcal A}_{\jpsi K^{\pm}} = (\,0.3\,\pm 3.0 \, \pm 0.4\,) \% \
.
\end{eqnarray}
Both results are consistent with no direct \CP\ violation in $B
\to \jpsi K$ decays.

\section{Could $\stwob_{\jpsi \KS}$ and $\stwob_{\jpsi \KL}$
differ?}

The main final states for the measurement of \stwob\ are $\jpsi
\KS$ and $\jpsi \KL$. Since the two measurements are combined, one
may ask the question: How good is the relation
\begin{eqnarray}
\label{eq:jpsikskl}
 \stwob_{\jpsi \KS} = \stwob_{\jpsi \KL} \ \ ?
\end{eqnarray}
 The interference between $\Bz \to
\jpsi \Kz$ and $\Bz \to \Bzb \to \jpsi \Kzb$ is possible thanks to
\KzKzb\ mixing. Obviously relation~(\ref{eq:jpsikskl}) cannot be
rigourously exact since the \KS\ and \KL\ physical states are not
exactly \CP\ eigenstates of the neutral kaon system.  However the
effect of indirect \CP\ violation in \KzKzb\ mixing has negligible
impact on~(\ref{eq:jpsikskl}). It has been shown~\cite{grossman}
that the only effect that could spoil this relation would be a
sizable wrong-flavor $\B \to \jpsi \Kb$ transition, which is not
possible at lowest orders in the Standard Model.  To investigate
the possibility of wrong-flavor decays, we study time-dependent
rates on samples of 860 (resp. 856) self-tagged $\jpsi K^{\ast 0}$
(resp. $\jpsi \Kb^{\ast 0}$) candidates (selected with a purity
greater than 96\%). We obtain the preliminary measurements
\begin{eqnarray}
\Gamma( \Bzb \to \jpsi \Kstarz ) / \Gamma( \Bz \to \jpsi \Kstarz )
&=& -0.022 \pm 0.028 \pm 0.016 \ , \nonumber \\
\Gamma( \Bz \to \jpsi \Kstarzb ) / \Gamma( \Bzb \to \jpsi \Kstarzb
) &=& +0.017 \pm 0.026 \pm 0.016 \ ,
\end{eqnarray}
which are consistent with the absence of wrong-flavor $\bar{b} \to
c \bar{c} s$ transitions, as expected.

\section{Search for \CP\ and \T\ violation and test of \CPT\
symmetry in \BzBzb\ mixing; limit on the lifetime difference.}

\par
The  differential rates of $\FourS \to \BB \to f_{\rm rec} f_{\rm
tag}$  events as a function of $\Delta t$ has an exponential
dependance $e^{-\Gamma |\Delta t|} $ modulated by a cosine and a
sine term at the mixing frequency $\Delta m$.  The exponential
envelop is slightly modified by terms that depend on the lifetime
difference $\Delta \Gamma$, while the coefficients of the cosine
and sine terms receive small corrections that depend on the \CP-
and \CPT-violating parameter $z$.  From a simultaneous
time-dependent fit to the \CP\ and flavor eigenstate samples
including tagged and untagged events, we obtain the following
preliminary measurements~\cite{babar_cpt}:
\begin{eqnarray}
\begin{array}{rcrccclc}
  {\rm sign}( {\rm Re}\, \lambda_{\CP}) \times \Delta \Gamma /  \Gamma &                        = & -0.008 & \pm & 0.037  & \pm & 0.018  & \ [\,-0.084,+0.068\,] \\
   \left| \, q / p \, \right|  &                                                          = &  1.029 & \pm & 0.013  & \pm & 0.011  & \ [\,+1.001,+1.057\,] \\
  \left(\,{ {\rm Re}\, \lambda_{\CP}}/\, {| \, \lambda_{\CP}  \, | }\, \right) \times {\rm Re}\, z  & = &  0.014 & \pm & 0.035  & \pm & 0.034  & \ [\,-0.072,+0.101\,] \\
   {\rm Im}\, z  &                                                                        = &  0.038 & \pm & 0.029  & \pm & 0.025  & \ [\,-0.028,+0.104\,] \\
\end{array}
\end{eqnarray}
where the first errors are statistical, the second errors
systematical, and 90\% confidence level intervals are given under
brackets.  The first parameter, $\Delta \Gamma /\Gamma$ with a
sign ambiguity, is consistent with zero within 4\%.  The second
parameter, which measures \CP\ and \T\ violation in mixing, is
consistent with unity within two standard deviations.  The last
two parameters, which are \CPT-violating, are consistent with
zero.
\par
Systematics include possible biases due to charge asymmetries in
the detector, which are estimated from the data, and a component
that covers  any effect due to quantum interference between the
two decays when the $B_{\rm tag}$ undergoes a doubly-Cabibbo
suppressed transition. Extensive systematic cross-checks include
alternative measurements of \stwob\ and $\Delta m$, which are
fully consistent with \babar\ published values and world averages.
We also confirm that the ratio $|\overline{A}_{\CP}/A_{\CP}|$ is
consistent with unity (no direct \CP\ violation in $\jpsi K$)
within 4.5\%.

\section{Measurements of \stwob\ using non-$\bar{b} \to \bar{c} c \bar{s}$
modes}

One of the most promising ways to look for New Physics at \B\
factories is to measure the \CP\ parameter \stwob\ in several \B\
decay modes sensitive to different short-distance physics.

The $ B \to \phi K$ decay is dominated by a pure $\bar{b} \to
\bar{s} s \bar{s}$ penguin transition.  Our updated preliminary
branching fraction and charge asymmetries measurements in these
modes are
\begin{eqnarray}
{\mathcal B}( \Bz \to \phi \Kz ) &=& (\,7.6\,^{+1.3}_{-1.2} \, \pm
0.5\,) \times 10^{-6} \nonumber \\
 {\mathcal B}( B^{+} \to \phi K^{+} ) &=&
(\,10.0\,^{+0.9}_{-0.8} \, \pm 0.5\,) \times 10^{-6} \\
{\mathcal A}_{\CP}( B^{\pm} \to \phi K^{\pm} ) &=& (\,3.9\,\pm 8.6
\, \pm 1.1\,) \% \nonumber
\end{eqnarray}
and we place a limit on the $B^+ \to \phi \pi^+$ decay
\begin{eqnarray}
{\mathcal B}( B^{+} \to \phi \pi^{+} ) < 0.38 \times 10^{-6} \ \ @
\, 90\%\, {\rm CL}
\end{eqnarray}
which indicates that the magnitude of rescattering in the $\phi K$
final states is small~\cite{babar_phik}. The $B$ decay final state
$\phi \KS$ is a \CP-odd eigenstate. Any deviation from $S_{\phi
\KS} = \stwob$ would be a strong indication of New Physics.  We
update our preliminary result~\cite{babar_phikscp} with a larger
data set (84 million \BB\ pairs) and augment it with the $\KS \to
\pi^0\pi^0$ channel. Based on a sample of $51.5\pm7.5$ candidates
in the $\KS \to \pi^+\pi^-$ channel and $13.3 \pm 5.3$ candidates
in the $\KS \to \pi^0 \pi^0$ channel we obtain:
\begin{eqnarray}
S_{\phi\KS} =  -0.18 \pm 0.51 \pm 0.07 \ ,\ \  C_{\phi\KS} =
-0.80 \pm 0.38 \pm 0.12 \ .
\end{eqnarray}
The large value of $C_{\phi \KS}$ reflects the mismatch between
$~27$ \Bz-tagged and $~13$ \Bzb-tagged events in our signal
sample. Fixing $C_{\phi \KS}$ to zero, we obtain $ S_{\phi\KS} =
-0.26 \pm 0.51 {(\rm stat)}$.

The $B \to \eta^{\prime} K$ decay is also a dominantly $\bar{b}
\to \bar{s} s \bar{s}$ transition.  However, because the
$\eta^{\prime}$ meson has non-negligible $\bar{u} u$ component,
the decay may also receive an additional $\bar{b} \to \bar{u} u
\bar{s}$ contribution with a different weak phase.  The
predictions are that the size of this non-penguin contribution is
relatively small.  We measure the branching fractions and charge
asymmetry~\cite{babar_etapk}
\begin{eqnarray}
{\mathcal B}( \Bz \to \eta^{\prime} \Kz ) &=& (\,55.4\,\pm5.2, \pm
4.0\,) \times 10^{-6} \nonumber \\
{\mathcal B}( B^{+} \to \eta^{\prime} K^{+} ) &=&
(\,76.9\,\pm 3.5 \, \pm 4.4\,) \times 10^{-6} \\
{\mathcal A}_{\CP}( B^{\pm} \to \eta^{\prime} K^{\pm} ) &=&
(\,3.7\,\pm 4.5 \, \pm 1.1\,) \% \nonumber
\end{eqnarray}
and time-dependent \CP\ parameters
\begin{eqnarray}
S_{\eta^{\prime}\KS} = +0.02 \pm 0.34 \pm 0.03 \ ,\ \
C_{\eta^{\prime}\KS} = +0.10 \pm 0.22 \pm 0.03 \ .
\end{eqnarray}
Provided that the tree contribution is small,
$S_{\eta^{\prime}\KS}$ is equal to \stwob.

The $B \to D^{(\ast)-} D^{(\ast)+}$ modes receive contribution
from Cabibbo-suppressed $\bar{b} \to \bar{c} c \bar{d}$ tree and
$\bar{b} \to \bar{d}$ penguin amplitudes. The latter is believed
to be much smaller than the former.  At present only the $\Bz \to
D^{\ast-} D^{\ast +}$ and the $\Bz \to D^{\ast \mp} D^{\pm}$ have
been observed. Using a sample of $126 \pm 13$ events selected in
the pseudoscalar to vector-vector  $\Bz \to D^{\ast +} D^{\ast -}$
mode, we performed a transversity analysis to disentangle the
$\CP=+1$ and $\CP=-1$ components of this decay.  As
anticipated~\cite{rosner} we find that the decay proceeds mostly
through the \CP-even component: $R_\perp = 0.07 \pm 0.06 \pm
0.03$. If the penguin contribution can be neglected, the imaginary
part of the \CP\ parameter $(\lambda_{D^{\ast}D^{\ast}})_{\CP=+1}$
is equal to $-\stwob$.  We obtain~\cite{babar_dstdst}:
\begin{eqnarray}
|(\lambda_{D^{\ast}D^{\ast}})_{\CP=+1}| = 0.98 \pm 0.25 \pm 0.13 \
\ \ {\rm and } \ \ \ {\rm Im} \,
(\lambda_{D^{\ast}D^{\ast}})_{\CP=+1} = 0.31 \pm 0.43 \pm 0.13 \ .
\end{eqnarray}

From a sample of $113 \pm 13$ events reconstructed in the modes
$D^{\ast -} D^+$ and $D^{\ast +} D^-$, we
measure~\cite{babar_dstd}:
\begin{eqnarray}
{\mathcal B}( \Bz \to D^{\ast\pm}D^\mp ) = (\,8.8\,\pm1.0, \pm
1.3\,) \times 10^{-4} \ { \rm and} \  {\mathcal A}_{D^{\ast}D} =
(\,-3\,\pm 11 \, \pm 5\,) \% \ ,
\end{eqnarray}
where $ {\mathcal A}_{D^{\ast}D} \equiv ( {\mathcal N}_{D^{\ast +}
D^-}-{\mathcal N}_{D^{\ast -} D^+} )/( {\mathcal N}_{D^{\ast +}
D^-}+{\mathcal N}_{D^{\ast -} D^+} ) $ is the time-integrated
charge asymmetry.  Our preliminary time-dependent results in these
modes are:
\begin{eqnarray}
S_{D^{\ast -} D^+} = -0.24 \pm 0.69 \pm 0.12 \ ,\ \ C_{D^{\ast -}
D^+} = -0.22 \pm 0.37 \pm 0.10 \ , \nonumber \\
S_{D^{\ast +} D^-} = -0.82 \pm 0.75 \pm 0.14 \ ,\ \ C_{D^{\ast +}
D^-} = -0.47 \pm 0.40 \pm 0.12 \ ,
\end{eqnarray}
where $ S_{D^{\ast -} D^+} $ and $ S_{D^{\ast +} D^-} $ are equal
to $ - \stwob$ if penguin contributions are negligible.

The $\Bz \to \jpsi \pi^0$ decay is a $\bar{b} \to \bar{c} c
\bar{d}$ transition. The dominant color-suppressed tree amplitude
is also Cabibbo-suppressed. The sub-dominant penguin amplitude is
CKM-suppressed but has a different weak phase, which might spoil
the relation $S_{\jpsi \pi^0} = -\stwob$.  With a sample of
$40\pm7$ signal events, we obtain~\cite{babar_jpsipi}:
\begin{eqnarray}
S_{\jpsi \pi^0} = +0.05 \pm 0.49 \pm 0.16 \ ,\ \ C_{\jpsi \pi^0} =
+0.38 \pm 0.41 \pm 0.09 \ .
\end{eqnarray}

The time-dependent results presented here are combined in
Table~\ref{tab:average} with corresponding results from Belle when
available. More statistics is needed to draw conclusions on the
slight discrepancy between the values of \stwob\ measured in
$(\bar{c} c) K$ and $(\bar{s} s) K$ modes.

\begin{table}[t]
\caption{Summary of time-dependent measurements in various modes
measuring \stwob, from the Heavy Flavor Averaging Group. The sign
of the $S$ coefficients for the $\jpsi \pi^0$ and $D^{\ast
\mp}D^{\pm}$ modes are reversed to be directly comparable to the
reference value of \stwob. \label{tab:average} } \vspace{0.4cm}
\begin{center}
\begin{tabular}{|l|c|c|c|l|  }
\hline
 mode $f$ &  $- \eta_f \times S_f$ ("\stwob") & $C_f$ &  transition & references \\
 \hline
 Charmonium \KS\ &  $ +0.734 \pm 0.055 $ & $ +0.052\pm0.047 $ & $ \bar{b} \to \bar{c} c \bar{s} $ & \babar~\cite{babar_stwob}, Belle~\cite{belle_stwob}    \\
 \hline
 $\phi \KS$ &  $ -0.38 \pm 0.41 $ & $ -0.19 \pm 0.30 $ & $ \bar{b} \to \bar{s} s \bar{s}$  & \babar~\cite{babar_gautier}, Belle~\cite{belle_sqq} \\
 $\eta^{\prime} \KS$ &  $ +0.33 \pm 0.25 $ & $ -0.08 \pm 0.16 $ &   & \babar~\cite{babar_etapk}, Belle~\cite{belle_sqq} \\
 \hline
$ D^{\ast-}D^{\ast+}$ &  $ -0.32 \pm 0.45 $ & $ +0.02 \pm 0.27 $ & $ \bar{b} \to \bar{c} c \bar{d}$ & \babar~\cite{babar_dstdst}, Belle   \\
$ D^{\ast-}D^{+} $ &  $ +0.24 \pm 0.70 $ & $ -0.22 \pm 0.38 $ &    & \babar~\cite{babar_dstd}  \\
$ D^{\ast+}D^{-} $ &  $ +0.82 \pm 0.76 $ & $ -0.47 \pm 0.42 $ &    &  \\
\hline
$\jpsi \pi^0 $ &  $ +0.47 \pm 0.41 $ & $ +0.26 \pm 0.29 $ & $ \bar{b} \to \bar{c} c \bar{d}$ & \babar~\cite{babar_jpsipi}, Belle~\cite{belle_jpsipi}   \\
 \hline
 \end{tabular}
\end{center}
\end{table}

\section{Time-dependent analyses in charmless modes}

The $B \to \pi^+ \pi^-$ decay receives contributions from
CKM-suppressed $\bar{b} \to \bar{u}$ tree and $ \bar{b} \to
\bar{d} $ penguin amplitudes. In the absence of penguin
contribution the phase of $\lambda_{\pi \pi }$ is
$-2(\beta+\gamma)$, which is equivalent to $2\alpha$ using the
triangle relation $\alpha + \beta + \gamma = \pi$ (here, $\gamma$
is the phase of ${V_{ub}}^{\ast}$). In presence of penguin
contribution  the modulus and phase of $\lambda_{\pi \pi }$ are
modified: $\lambda_{\pi \pi} \equiv |\lambda_{\pi \pi}|\,
e^{2i\alpha_{\rm eff}}$.  The observables are $C_{\pi \pi} =
(1-|\lambda_{\pi \pi}|^2)/(1+|\lambda_{\pi \pi}|^2)$ and $S_{\pi
\pi} = \sqrt{ 1 - C_{\pi \pi}^{ \, 2} } \, \sin{ 2\alpha_{\rm
eff}}$.  One condition for direct \CP\ violation ($C_{\pi \pi}
\neq 0$) is that the relative strong phase $\delta_{\pi \pi}$
between the tree and penguin amplitudes be non-zero, while $\alpha
- \alpha_{\rm eff}$ depends on the absolute ratio $|P/T|$ of the
penguin to tree amplitude.

We perform a simultaneous $\pi^+ \pi^-/K^+ \pi^-$ analysis. The
Cherenkov angles as measured in the DIRC enter directly the
likelihood function as discriminating variables to distinguish
between the $\pi^+ \pi^-$ and $K^+ \pi^-$ modes.  The separation
that results from this discrimination is excellent. The maximum
likelihood fit identifies ${\mathcal N}_{K\pi} = 589 \pm 30$ and
${\mathcal N}_{\pi \pi} = 157 \pm 7$ signal candidates out of a
large continuum-dominated sample of events.  From the self-tagged
$K \pi$ sample we measure $\Delta m$ as a cross-check and find a
value in full agreement with the actual value.  From the $\pi \pi$
sample we measure~\cite{babar_pipi}:
\begin{eqnarray}
S_{\pi \pi} = +0.02 \pm 0.34 \pm 0.05 \ ,\ \ C_{\pi \pi} = -0.30
\pm 0.25 \pm 0.04 \ .
\end{eqnarray}
The statistical errors are quoted from the likelihood fit and are
in good agreement with expectation from Monte-Carlo studies.  Our
result is well into the physical region ($C_{\pi \pi}^{\, 2} +
S_{\pi \pi}^{\, 2} \leq 1$) and does not bring evidence for either
direct or mixing-induced \CP\ violation in the $B \to \pi^+ \pi^-$
mode: \babar\ does not confirm the observation by Belle of large
\CP\ violation effects in this mode~\cite{belle_pipi}.

The time-dependent study of the $B \to \pi^+ \pi^- \pi^0$ decay is
in principle a promising way for a model-independent determination
of angle $\alpha$~\cite{quinn}. In practice the description of the
interfering resonant structure in the $3\pi$ Dalitz plot will
introduce some level of model dependence in the extraction of
$\alpha$, which has to be evaluated.  With the present statistics
we perform a quasi two-body analysis where we select bands around
the $\rho^+$ and $\rho^-$ in the $\pi^+ \pi^- \pi^0$ Dalitz plot,
excluding the interfering region at the intersection between the
two bands, where the charge assignment ($\rho^+ \pi^-$ or $\rho^-
\pi^+$) is ambiguous.  The analysis is similar to that in the
$\pi^+\pi^-$ mode, with the additional complications of a $\pi^0$
in the final state and of a larger background from poorly-known
rare \B\ decays.  We perform a simultaneous $\rho \pi/ \rho K$
analysis that yields ${\mathcal N}_{\rho \pi} = 428 \pm 42$ and
${\mathcal N}_{\rho K} = 120 \pm 28$ signal events.  The small
ratio $\rho K / \rho \pi $ is an indication that the penguin
contribution is smaller in the $\rho \pi$ mode than it is in the
$\pi \pi$ mode, as anticipated. We obtain the preliminary
branching fractions and charge asymmetry:
\begin{eqnarray}
{\mathcal B}( \Bz \to \rho^\pm \pi^\mp ) &=& (\,22.6\, \pm 1.8 \,
\pm 2.2\,) \times 10^{-6}  \nonumber \\
{\mathcal B}( \Bz \to \rho^\pm K^\mp ) &=&
(\ 7.3\,^{+1.3}_{-1.2} \, \pm 1.3\,) \times 10^{-6} \\
{\mathcal A}_{\rho K} &=& (\,+28\,\pm 17 \, \pm 8\,) \%  \ .
\nonumber
\end{eqnarray}
From the $\rho \pi$ sample we update our time-dependent \CP\
measurements~\cite{babar_rhopi} in this mode.  In addition to a
global charge asymmetry ${\mathcal A}_{\rho \pi} = (\,+18 \, \pm 8
\, \pm 3\,) \%$, we obtain:
\begin{eqnarray}
S_{\rho \pi} = +0.19 \pm 0.24 \pm 0.03\,  , \ C_{\rho \pi} = +0.36
\pm 0.18 \pm 0.04 \, ,  \nonumber \\
\Delta S_{\rho \pi} = +0.15 \pm 0.25 \pm 0.03\, , \ \Delta C_{\rho
\pi} = +0.28 \pm 0.19 \pm 0.04\,  .
\end{eqnarray}
$S_{\rho \pi}$ and $C_{\rho \pi}$ are parameters that measure
mixing-induced and direct \CP\ violation, respectively. $\Delta
C_{\rho \pi}$ and $\Delta S_{\rho \pi}$ are dilution parameters:
$\Delta C_{\rho \pi}$ is linked to the ratio of $\Bz \to \rho^-
\pi^+$ and $\Bz \to \rho^+ \pi^-$ amplitudes, and its value is
consistent with predictions; $\Delta S_{\rho \pi}$ is a
non-trivial combination of strong and weak phase differences.  The
values of ${\mathcal A}_{\rho \pi}$ and $C_{\rho \pi}$ can be
interpreted as a $\sim 2.5 \sigma $  deviation from the hypothesis
of no-direct \CP\ violation in this mode, from which no claim can
be made.

\section{Conclusions}

A broad experimental program of time-dependent measurements in a
variety of modes related to angles $\beta$ and $\alpha$ of the
Unitarity Triangle is underway at \babar.  With the exception of
the main \stwob\ measurement in golden channels, these
measurements have poor statistical significance, but promise
exciting results with an order-of-magnitude larger statistics at
\B\ factories.  Eventually this array of measurements will put
strong constraints on the fundamental parameters of the CKM model
in the Standard Model and, perhaps, reveal the presence of New
Physics in processes involving \B\ meson mixing and decay.

\section{Acknowledgements}
I am grateful to my \babar\ colleagues for their help and support
in preparing this talk. I would like to congratulate the
organizers for the outstanding scientific quality of the
conference.

\end{document}